\documentclass{emulateapj}
\input{epsf}
\begin{document}

\title{Why do the braking indices of pulsars span a range of more than 100 millions?}
%\begin{CJK*}{GB}{gbsn}

%\author{Shuang-Nan Zhang ( уек╚до )\altaffilmark{1,2}, Yi Xie\altaffilmark{1}}
\author{Shuang-Nan Zhang\altaffilmark{1,2}, and Yi Xie\altaffilmark{1}}
\altaffiltext{1}{National Astronomical Observatories, Chinese Academy of Sciences, Beijing, 100012, China;
zhangsn@ihep.ac.cn}\altaffiltext{2}{Key Laboratory of Particle Astrophysics, Institute of High Energy Physics,
Chinese Academy of Sciences, Beijing 100049, China}
%\end{CJK*}

\begin{abstract}

Here we report that the observed braking indices of the 366 pulsars in the sample of Hobbs et al. range from
about $-10^8$ to about $+10^8$ and are significantly correlated with their characteristic ages. Using the model
of magnetic field evolution we developed previously based on the same data, we derived an analytical expression
for the braking index, which agrees with all the observed statistical properties of the braking indices of the
pulsars in the sample of Hobbs et al. Our model is, however, incompatible with the previous interpretation that
magnetic field growth is responsible for the small values of braking indices ($<3$) observed for ``baby" pulsars
with characteristic ages of less than $2\times 10^3$~yr. We find that the ``instantaneous" braking index of a
pulsar may be different from the ``averaged" braking index obtained from fitting the data over a certain time
span. The close match between our model-predicted ``instantaneous" braking indices and the observed ``averaged"
braking indices suggests that the time spans used previously are usually smaller than or comparable to their
magnetic field oscillation periods. Our model can be tested with the existing data, by calculating the braking
index as a function of the time span for each pulsar. In doing so, one can obtain for each pulsar all the
parameters in our magnetic field evolution model, and may be able to improve the sensitivity of using pulsars to
detect gravitational waves.

\end{abstract}

\keywords{magnetic fields - methods: statistical - pulsars: general - stars: neutron}

%%%%%%%%%%%%%%%%%%%%%%%%%%%%%%%%%%%%%%%%%%%%%%%%%%%%%%%%%%%%%%%%
\section{Introduction}           %% first-level sections will be auto-capitalized

Assuming the braking law of a pulsar as
\begin{equation}
\dot \nu  =-K\nu ^{n_{\rm b}},\label{braking_law}
\end{equation}
where $\nu$ is its spin frequency and $n_{\rm b}$ is called its braking index, Manchester \& Taylor (1977) found
that
\begin{equation}
n_{\rm b}=\ddot{\nu}\nu/\dot{\nu}^2,\label{braking_index}
\end{equation}
if $\dot{K}=0$. For the standard magnetic dipole radiation model with constant magnetic field ($\dot{K}=0$),
$n_{\rm b}=3$. Therefore $n_{\rm b}\ne 3$ indicates some deviation from the standard magnetic dipole radiation
model with constant magnetic field.

Blandford \& Romani (1988) re-formulated the braking law of a pulsar as,
\begin{equation}
\dot \nu  =-K(t)\nu ^3.\label{blandford1}
\end{equation}
This means that the standard magnetic dipole radiation is responsible for the instantaneous spin-down of a
pulsar, but the braking torque determined by $K(t)$ may be variable. In this formulation, $n_{\rm b}\ne 3$ does
not indicate deviation from the standard magnetic dipole radiation model, but means only that $K(t)$ is time
dependent. From Equation~(\ref{blandford1}) one can obtain,
\begin{equation}
\frac{\dot{K}}{K}\frac{\nu}{\dot{\nu}}={n_{\rm b}}-3.\label{blandford2}
\end{equation}
Assuming that magnetic field evolution is responsible for the variation of $K(t)$, we have $K=AB(t)^2$, in which
$A$ is a constant and $B(t)$ is the time variable dipole magnetic field strength of a pulsar. The above equation
then suggests that $n_{\rm b}< 3$ indicates magnetic field growth of a pulsar, and vice versa, since
$\dot{\nu}<0$ and $K>0$. This can be seen more clearly from (Zhang \& Xie 2012b; hereafter Paper I),
\begin{equation}
\dot{K}=\frac{\ddot{\nu}}{\nu^3}(\frac{3}{n_{\rm b}}-1)=\frac{\dot{\nu}^2}{\nu^4}(3-n_{\rm b}).\label{zhang2012}
\end{equation}

Many authors have thus used the observed $n_{\rm b}<3$ of some very young pulsars to infer their possible
magnetic field increase (e.g., Blandford \& Romani 1988; Chanmugam \& Sang 1989; Manchester \& Peterson 1989;
Lyne et al. 1993, 1996; Johnston \& Galloway 1999; Wu et al. 2003; Lyne 2004; Chen \& Li 2006; Livingstone et
al. 2006, 2007; Middleditch et al. 2006; Yue et al. 2007; Espinoza et al. 2011). Magnetospheric activities or
fall-back accretion may also be able to produce $n_{\rm b}<3$ (e.g. Menou et al. 2001; Alpar et al. 2001; Xu \&
Qiao 2001; Wu et al. 2003; Chen \& Li 2006; Yue et al. 2007). This can also be explained by another model based
on a decrease in effective moment of inertia due to an increase in the fraction of the stellar core that becomes
superfluid as the star cools via neutrino emission (Ho \& Andersson 2012).

However, the above studies have not investigated the statistical properties of the braking indices of pulsars.
Recently, Magalhaes et al. (2012) studied the observed braking indices of several very young pulsars
statistically. They built a model of a pulsar's braking law to explain the ranges of these observed braking
indices, and then predicted the possible values of the braking indices of several other very young pulsars
withusing only their measured $\nu$ and $\dot\nu$. Testing these predictions and applying the model to more
pulsars for observational tests may shed light on our further understanding of pulsars.

In Paper I, we have shown that the observed $n_{\rm b}$, for the large sample of pulsars reported by Hobbs et
al. (2010, hereafter H2010), not only deviates from $n_{\rm b}=3$, but also ranges from around $-10^8$ to around
$+10^8$, i.e., spanning a range of more than 100 millions, which has not been explained so far. In this work, we
will first apply the model of Magalhaes et al. (2012) to this sample, and then search for possible correlations
of the observed $n_{\rm b}$ of pulsars in the sample of H2010. We find a significant correlation between $n_{\rm
b}$ and the characteristic ages of pulsars, which is explained satisfactorily with the magnetic field evolution
model we developed in Paper I. Finally we will make a testable prediction of our model and discuss its physical
implications.

\section{Applicability of the model of Magalhaes et al. (2012) to the pulsars in H2010}

Magalhaes et al. (2012) used equation~({\ref{braking_law}) to predict,
\begin{equation}
n_{\rm pre}=\frac{\log(|\dot{\Omega}|S^2/\xi^2)}{\log(\Omega)},\label{m2012}
\end{equation}
where $\Omega=2\pi \nu$, $S=2.3\times 10^{20}$~Hz$^{1/2}$~G and is assumed to be the same for all pulsars, and
$1.5\times 10^{13}<\xi<34\times 10^{13}$ in units of Hz$^{(3-n_{\rm b})/2}$~G determines the range of $n_{\rm
pre}$ for each pulsar. Magalhaes et al. (2012) showed that Equation~(\ref{m2012}) describes the ranges of
$n_{\rm b}$ observed for several very young pulsars very well. A natural question arises: can this model also
explain the observed $n_{\rm b}$ for other older pulsars of the sample of H2010? In Figure~\ref{Fig:1} we plot
the distributions of the observed braking indices ($n_{\rm obs}=\ddot{\nu}\nu/\dot{\nu}^2$) of the pulsars of
the sample of H2010 and the ratios between $n_{\rm obs}$ and $n_{\rm pre}$. $n_{\rm obs}$ spans a range from
around $-10^8$ to around $+10^8$, as shown in Paper I. However, $n_{\rm obs}/n_{\rm pre}$ spans a similar range
and has almost the same numbers of positive and negative values. This means that the model of Magalhaes et al.
(2012) cannot explain both the magnitudes and signs of the braking indices of the pulsars of the sample of
H2010. Therefore an alternative model is needed to account for the observed braking indices of these pulsars.
\begin{figure}
\center{
\includegraphics[angle=0,scale=0.40]{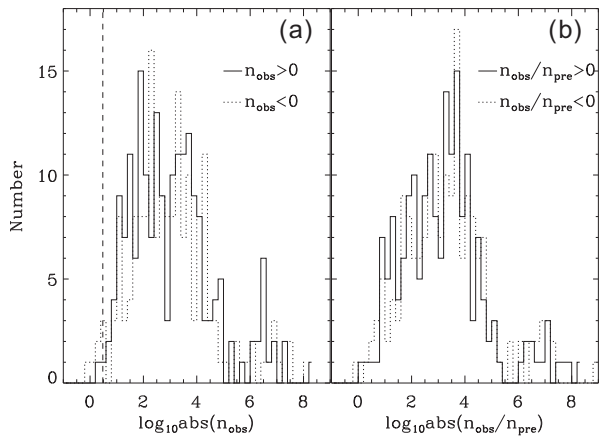}}
\caption{(a) Distributions of the  observed braking indices ($n_{\rm obs}=\ddot{\nu}\nu/\dot{\nu}^2$). (b)
Ratios of the observed and predicted braking indices ($n_{\rm pre}$) from Equation~(\ref{m2012}) (equation ~(11)
in Magalhaes et al. 2012).} \label{Fig:1}
\end{figure}

\section{Observed correlations of braking indices}

In order to develop a model to account for the observed braking indices of the pulsars of the sample of H2010,
we first explore possible correlations of the observed braking index $n_{\rm b}=\ddot{\nu}\nu/\dot{\nu}^2$ with
various spin-down parameters and their combinations, as shown in Figure~\ref{Fig:2}.  No or only weak
correlation is found either between $n_{\rm b}$ and $\nu$ or between $n_{\rm b}$ and $\ddot{\nu}$. Strong
correlations are found both between $n_{\rm b}$ and $\dot{\nu}$ and between $n_{\rm b}$ and the characteristic
age $\tau_{\rm c}=-\frac{\nu}{2\dot{\nu}}$; the latter correlation is very significant and is the focus of study
in the rest of this work.

The $n_{\rm b}\sim \tau_{\rm c}$ correlation shown in the bottom panel of Figure~\ref{Fig:2} reveals three
phenomena. (1) For all young pulsars with $10^{4}<\tau_{\rm c}<5\times 10^{4}$~yr, $10<n_{\rm b}<40$ and $n_{\rm
b}$ is not correlated with $\tau_{\rm c}$. (2) Excluding these pulsars, the overall correlation is almost linear
over about eight orders of magnitude for both $n_{\rm b}$ and $\tau_{\rm c}$. (3) The overall correlation is
almost the same for positive or negative $n_{\rm b}$.

\begin{figure}
\center{
\includegraphics[angle=0,scale=0.50]{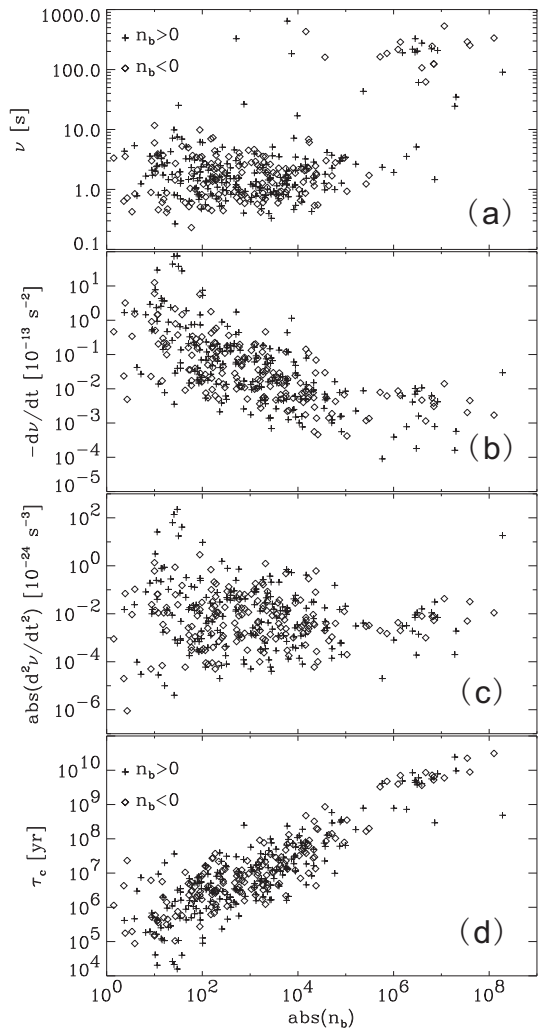}}
\caption{Observed correlations of the braking index $n_{\rm b}$ with (from top to bottom) $\nu$, $\dot \nu$,
$\ddot \nu$ and $\tau_{\rm c}$.} \label{Fig:2}
\end{figure}

\section{A new analytical model of braking index}

In Paper I, we have shown that the observed correlations between $\ddot \nu$ and other observables or parameters
can be well explained by assuming that the braking law takes the form of Equation~(\ref{blandford1}) and the
evolution of the dipole magnetic field of a pulsar consists of a long-term decay modulated by one or multiple
oscillation components, i.e.,
\begin{equation}
B(t)= B_0 (\frac{t}{t_{0}})^{-\alpha}(1+k\sin(\phi+2\pi\frac{t}{T})),\label{b_evol}
\end{equation}
where $t_{0}$ and $B_0$ are the initial decay time and dipole magnetic field strength, $\alpha$ is the decay
index, and $k$, $\phi$ and $T$ are the oscillation amplitude, initial oscillation phase, and oscillation period,
respectively. In Paper I we have shown that $\alpha\ge 0.5$, but its upper limit cannot be well constrained with
the existing data. In this work, we simply take $\alpha= 0.5$ for simplicity; the conclusions of this work will
not change for larger values of $\alpha$.

From Equation~(\ref{blandford1}), we can solve for the spin evolution of a pulsar,
\begin{equation}
\nu^{-2}=\nu_{0}^{-2}+2\int_{t_0}^{t}K(t)dt.\label{nu1}
\end{equation}
With $K=AB(t)^2$ and Equation~(\ref{b_evol}), we have
\begin{equation}
\tau_{\rm c}=-\frac{\nu}{2\dot{\nu}}=\frac{1}{2K\nu^2} \approx t\ln\frac{t}{t_0},\label{tau}
\end{equation}
where $\alpha=0.5$ and $k\ll 1$ are assumed. It is noticeable that the oscillation term in
Equation~(\ref{b_evol}) is not important in the above relation, since  $k\ll 1$. However the decay component of
Equation~(\ref{b_evol}) makes the real age of a pulsar significantly shorter than its characteristics age, since
normally $t/t_0\gg 1$ (Zhang \& Xie, 2012a).

From Equation~(\ref{nu1}) we can easily obtain $\dot{\nu}(t)$ and $\ddot{\nu}(t)$. Inserting $\nu(t)$,
$\dot{\nu}(t)$ and $\ddot{\nu}(t)$ into Equation~(\ref{braking_index}), we have,
\begin{equation}
n_{\rm b} =3+ \ln\frac{t}{t_0} (2- 4ft C(t))=3+\frac{\tau_{\rm c}}{t}(2- 4 ft C(t)),\label{n_b}
\end{equation}
where $f=\frac{2\pi k}{T}$ and $C(t)=\cos(\phi+2\pi \frac{t}{T})$. As we have shown in Paper I, the oscillatory
term can be ignored in determining $\ddot{\nu}$, so young pulsars with $\tau_{\rm c}\le 10^5$ have
$\ddot{\nu}>0$. Ignoring the oscillatory term in Equation~(\ref{n_b}), we have
\begin{equation}
n_{\rm b} \approx 3+2\frac{\tau_{\rm c}}{t}=3+2\ln\frac{t}{t_0},\label{n_y}
\end{equation}
which for young pulsars gives $n_{\rm b}>0$ (since $t>t_0$) and $n_{\rm b}$ is not well correlated with
$\tau_{\rm c}$, in agreement with the data shown in the lower left corner of the bottom panel in
Figure~\ref{Fig:2}. For old pulsars, the oscillatory term dominates and thus Equation~(\ref{n_b}) gives,
\begin{equation}
n_{\rm b}  \approx \pm 4\tau_{\rm c}f,
\end{equation}
since $C(t)$ most likely takes values around $+1$ or $-1$. This immediately suggests that $n_{\rm b}$ should be
linearly correlated with $\tau_{\rm c}$ and have equal probabilities to be either positive or negative, again in
agreement with the correlation shown in the bottom panel of Figure~\ref{Fig:2}.

In Figure~\ref{Fig:3}, we plot the analytically calculated $n_{\rm b}$ as a function of $\tau_{\rm c}$ using
Equations~(\ref{tau}) and (\ref{n_b}) for $k=10^{-4}$ and $10^{-5}$, respectively; we also assume $t_0=0.5$~yr
and $T=10$~yr, the latter is the typical time scale of the quasi-periodic features in the power spectra of the
timing residuals reported in H2010. The upper boundary in each panel is obtained when $C(t)=\pm 1$ and the other
values of $n_{\rm b}$ are obtained when $-1<C(t)<1$. Clearly the observed correlation shown in
Figure~\ref{Fig:2} is well reproduced. In particular, the lack of $n_{\rm b}<0$ for young pulsars with
$\tau_{\rm c}$ is predicted with $k\le 10^{-4}$, which in our model is caused by their long term magnetic field
decay.
\begin{figure*}
\center{
\includegraphics[angle=0,scale=0.50]{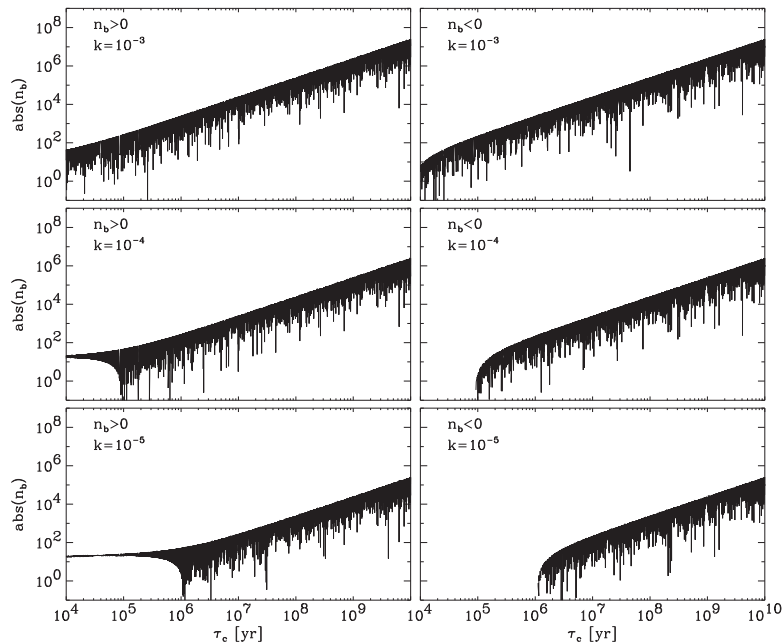}}
\caption{Analytically calculated $n_{\rm b}$ as a function of $\tau_{\rm c}$ using Equation~(\ref{n_b}), for
different values of $k$.}\label{Fig:3}
\end{figure*}

Equation~(\ref{n_b}) can be rewritten as,
\begin{equation}
n_{\rm b}/\tau_{\rm c}=3/\tau_{\rm c}+2/t- 4 f C(t) \approx 3/\tau_{\rm c}+2/t\pm 4 f.\label{n_b1}
\end{equation}
In Figure~\ref{Fig:4}, we show the observed correlation between $n_{\rm b}/\tau_{\rm c}$ and $\tau_{\rm c}$,
overplotted with the analytical results of Equation~(\ref{n_b1}) with $T=10$~yr and $k=10^{-3},\ 10^{-4}$ and
$10^{-5}$, respectively. Once again, the analytical results agree quite well with the data quite well.

\begin{figure*}
\center{
\includegraphics[angle=0,scale=0.50]{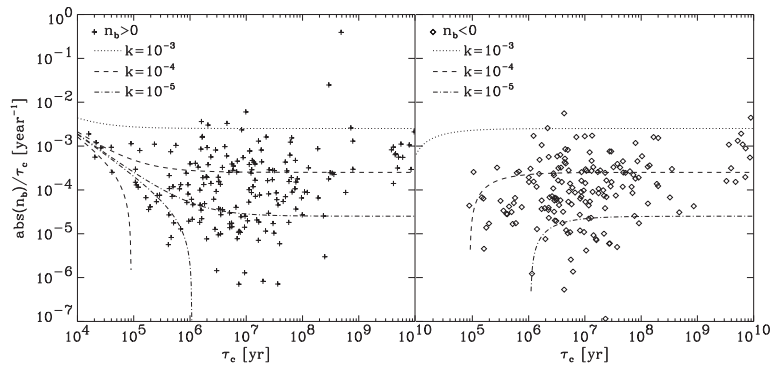}}
\caption{Correlation between $n_{\rm b}/\tau_{\rm c}$ and $\tau_{\rm c}$. The different curves show the
analytically calculated model predictions using Equation~(\ref{n_b1}) for different values of $k$.}
\label{Fig:4}
\end{figure*}

\section{On ``baby" pulsars and young pulsars}

As discussed above, $n_{\rm b}<3$ has been well established for all very young pulsars with $\tau_{\rm
c}<2\times 10^{3}$~yr (Livingstone et al. 2007), which we call ``baby" pulsars here for convenience. In
Figure~\ref{Fig:5} we show a comparison between all the measured braking indices of ``baby" pulsars and those of
young pulsars with $10^{4}<\tau_{\rm c}<10^{5}$~yr. From Equation~(\ref{zhang2012}), we get $\dot {K}>0$ or
$\dot {K}<0$ for ``baby" pulsars or young pulsars, respectively. If $\dot {K}\ne 0$ is due to $\dot {B}\ne 0$,
we are then led to the scenario that the magnetic fields of ``baby" pulsars grow, but those of young pulsars
decay. If these ``baby" pulsars with $\tau_{\rm c}<2\times 10^{3}$~yr evolve to become young pulsars, this must
happen when the time of transition from the growth phase to the decay phase is longer than the ages of these
``baby" pulsars, i.e., $t_0^{\rm baby}\geq 10^3$~yr. For a typical young pulsar with $\tau_{\rm c}\sim 2\times
10^{4}$~yr and $n_{\rm b}\sim 20$, Equation~(\ref{n_y}) gives $t\sim 2.4\times 10^3$ and $t_0^{\rm young}\sim
0.5$~yr. The mismatch between $t_0^{\rm baby}$ and $t_0^{\rm young}$ suggests that significantly different
mechanisms are responsible for the observed braking indices for ``baby" and young pulsars, respectively. In
other words, our model that $n_{\rm b}\ne 3$ is due to $\dot{B}\ne 0$ can only be applied to young (and old
pulsars), but not to ``baby" pulsars. Indeed, we may expect significant fall-back accretion or neutrino cooling
only for ``baby" pulsars; these mechanisms have previously also been suggested to be responsible for the
observed $n_{\rm b}<3$ of these ``baby" pulsars. In either case, the additional torque of accretion or the
reduced moment of inertia due to an increased fraction of superfluid stellar core can result in $n_{\rm b}<3$.
Stronger magnetospheric activities are also expected from ``baby" pulsars, whose magnetic fields should be
stronger than those of young pulsars.

\begin{figure}
\center{
\includegraphics[angle=0,scale=0.30]{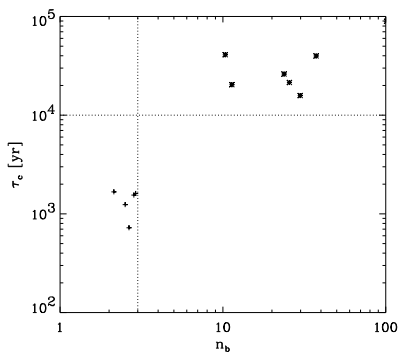}}
\caption{Comparison between the measured braking indices of ``baby" pulsars (in the lower left corner; data from
Livingstone et al. 2007) and those of young pulsars (in the upper right corner; data from H2010).} \label{Fig:5}
\end{figure}

\section{Model prediction: the effect of time span}
All the above equations for $n_{\rm b}$ give $n_{\rm b}$ as a function of $t$, i.e., the calculated $n_{\rm b}$
is in fact a function of time for a given pulsar, as shown in the left panel of Figure~\ref{Fig:6}, in which the
horizontal axis ``Time" is the calendar time. We call $n_{\rm b}$ calculated this way the ``instantaneous"
braking index. However, in analyzing the observed timing data of a pulsar, one usually fits the data on time of
arrival (TOA) over a certain time span to a Taylor expansion to third order:
\begin{equation}
\Phi (t) = \Phi_0 + \nu_0 (t-t_0) + \frac{1}{2}\dot \nu_0 (t-t_0)^2 + \frac{1}{6}\ddot\nu_0
(t-t_0)^3,\label{time_span}
\end{equation}
where $\Phi (t)$ is the phase of TOA of the observed pulses, and $\Phi_0$, $\nu_0$, $\dot \nu_0$ and
$\ddot\nu_0$ are the values of these parameters at $t_0$, to be determined from the fitting. $n_{\rm b}$
calculated from $\nu_0$, $\dot \nu_0$ and $\ddot\nu_0$ is thus not exactly the same as the ``instantaneous"
braking index. We call $n_{\rm b}$ calculated this way over a certain time span the ``averaged" braking index.

In the right panel of Figure~\ref{Fig:6}, we show the simulated result for the ``averaged" braking index as a
function of time span, In this simulation, we first produce a series of TOAs using Equations~(\ref{nu1}) and
(\ref{b_evol}), and then use Equation~(\ref{time_span}) to obtain $\nu_0$, $\dot \nu_0$ and $\ddot\nu_0$ for
different lengths of time span. It can be seen that the ``averaged" $n_{\rm b}$ is close to the ``instantaneous"
one when the time span is shorter than $T$, which is the oscillation period of the magnetic field. The close
match between our model predicted ``instantaneous" $n_{\rm b}$ and the observed ``averaged" $n_{\rm b}$, as
shown in Figure~\ref{Fig:4} suggests that the time spans used in the H2010 sample are usually smaller than $T$.

For some pulsars the observation history may be longer than $T$ and one can thus test the prediction of
Figure~\ref{Fig:6} with the existing data. In doing so, we can also obtain both $f$ and $T$ for a pulsar, thus
allowing a direct test of our model for a single pulsar. We can in principle then include the model of magnetic
field evolution for each pulsar in modeling its long term timing data, in order to remove the red noise in its
timing residuals, which may potentially be the limiting factor to the sensitivity in detecting gravitational
waves with pulsars.

\begin{figure}
\center{
\includegraphics[angle=0,scale=0.40]{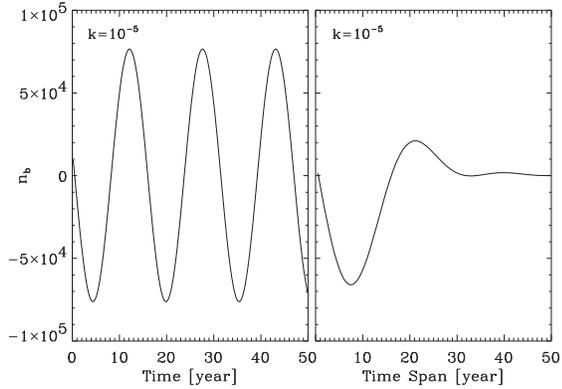}}
\caption{Left: ``instantaneous" braking index $n_{\rm b}$ as a function of time. Right: ``averaged" braking
index $n_{\rm b}$ as a function of the time span of the fitting. $T=15$~yr is used in both cases.} \label{Fig:6}
\end{figure}

\section{Summary}

The results and conclusions of this work can be summarized as follows.

\begin{enumerate}

\item The observed braking indices of the pulsars in the H2010 sample, which span a range of more than 100 millions,
are completely different from the predictions of the model of Magalhaes et al. (2012), as shown in
Figure~\ref{Fig:1}.

\item A significant correlation between the braking indices and characteristic ages of pulsars in the H2010
sample is found over about eight orders of magnitude, as shown in the bottom panel of Figure~\ref{Fig:2}.

\item Based on the magnetic field evolution model we developed previously (Paper I), an analytical expression
for
the braking index is derived, as given by Equation~(\ref{n_b}). The analytically calculated correlation between
the braking index and characteristic age re-produces the observed correlation well, as shown in
Figure~\ref{Fig:3}.

\item Our model is incompatible with the previous interpretation that the magnetic field growth is responsible for
the observed small values of braking indices ($<3$) of ``baby" pulsars with characteristic ages of less than
$2\times 10^3$~yr.

\item We find that the ``instantaneous" braking index of a pulsar may be different from the ``averaged"
braking index obtained from data. The close match between our model-predicted ``instantaneous" braking indices
and the observed ``averaged" braking indices, as shown in Figure~\ref{Fig:4} suggests that the time spans used
in the H2010 sample are usually smaller than or comparable to their magnetic field oscillation periods.

\item Our model can be tested with the existing data, by calculating the braking index as a function of the time
span for each pulsar. In doing so, one can obtain all the parameters for each pulsar in our magnetic field
evolution model. This is particularly important if one wants to use the long term timing data of pulsars to
detect gravitational waves.

\end{enumerate}

 \acknowledgments
The anonymous referee is thanked for a valuable suggestion which helped to clarify one important point in the
revised manuscript. S.-N.Z. acknowledges partial funding support by 973 Program of China under grant
2009CB824800, and by the National Natural Science Foundation of China under grant Nos. 11133002, 10821061, and
10725313.

{\it Note added}. After our Paper I was published and this manuscript was submitted, we noticed that Pons et al.
(2012) also proposed a phenomenologically similar model of magnetic field oscillations, which was also used to
explain the observed braking indices of older pulsars. In their model the oscillations are identified as due to
the Hall drift effect and have time scales of $(10^{6}$-$10^{8})/B$~yr, where $B$, in units of $10^{12}$~G, is
the surface dipole magnetic field strength of a pulsar. These time scales are much longer than those of
10-100~yr in our model. The proposed test of our model with Figure~\ref{Fig:6} can be used to distinguish
between their model and ours.


\begin{thebibliography}{}
\bibitem[Alpar et al.(2001)]{2001ApJ...557L..61A} Alpar, M.~A., Ankay, A., \& Yazgan, E.\ 2001, \apjl, 557, L61
\bibitem[Blandford \& Romani(1988)]{1988MNRAS.234P..57B} Blandford, R.~D., \& Romani, R.~W.\ 1988, \mnras, 234, 57P
\bibitem[Chanmugam \& Sang(1989)]{1989MNRAS.241..295C} Chanmugam, G., \& Sang, Y.\ 1989, \mnras, 241, 295
\bibitem[Chen \& Li(2006)]{2006A&A...450L...1C} Chen, W.~C., \& Li, X.~D.\ 2006, \aap, 450, L1
\bibitem[Espinoza et al.(2011)]{2011ApJ...741L..13E} Espinoza, C.~M., Lyne, A.~G., Kramer, M., Manchester, R.~N., \& Kaspi, V.~M.\ 2011, \apjl, 741, L13
\bibitem[Ho \& Andersson(2012)]{2012arXiv1208.3201H} Ho, W.~C.~G., \& Andersson, N.\ 2012, arXiv:1208.3201
\bibitem[Hobbs et al.(2010)]{2010MNRAS.402.1027H} Hobbs, G., Lyne, A.~G., \& Kramer, M.\ 2010, \mnras, 402, 1027 (H2010)
\bibitem[Johnston \& Galloway(1999)]{1999MNRAS.306L..50J} Johnston, S., \& Galloway, D.\ 1999, \mnras, 306, L50
\bibitem[Livingstone et al.(2007)]{2007Ap&SS.308..317L} Livingstone, M.~A., Kaspi, V.~M., Gavriil, F.~P., et al.\ 2007, \apss, 308, 317
\bibitem[Livingstone et al.(2006)]{2006ApJ...647.1286L} Livingstone, M.~A., Kaspi, V.~M., Gotthelf, E.~V., \& Kuiper, L.\ 2006, \apj, 647, 1286
\bibitem[Lyne(2004)]{2004IAUS..218..257L} Lyne, A.~G.\ 2004, Young Neutron Stars and Their Environments, IAU Symposium no. 218,  (Eds., Camilo, F. \& Bryan M. Gaensler, B. M.), 218, 257
\bibitem[Lyne et al.(1993)]{1993MNRAS.265.1003L} Lyne, A.~G., Pritchard, R.~S., \& Graham-Smith, F.\ 1993, \mnras, 265, 1003
\bibitem[Lyne et al.(1996)]{1996Natur.381..497L} Lyne, A.~G., Pritchard, R.~S., Graham-Smith, F., \& Camilo, F.\ 1996, \nat, 381, 497
\bibitem[Magalhaes et al.(2012)]{2012ApJ...755...54M} Magalhaes, N.~S., Miranda, T.~A., \& Frajuca, C.\ 2012, \apj, 755, 54
\bibitem[Manchester \& Peterson(1989)]{1989ApJ...342L..23M} Manchester, R.~N., \& Peterson, B.~A.\ 1989, \apjl, 342, L23
\bibitem[Manchester \& Taylor(1977)]{1977puls.book.....M} Manchester, R.~N., \& Taylor, J.~H.\ 1977, in Pulsars, ed. R. N. Manchester \& J.~H. Taylor (San Francisco, CA: W.~H.~Freeman), 281
\bibitem[Menou et al.(2001)]{2001ApJ...554L..63M} Menou, K., Perna, R., \& Hernquist, L.\ 2001, \apjl, 554, L63
\bibitem[Middleditch et al.(2006)]{2006ApJ...652.1531M} Middleditch, J.,
Marshall, F.~E., Wang, Q.~D., Gotthelf, E.~V., \& Zhang, W.\ 2006, \apj, 652, 1531
\bibitem[Pons et al.(2012)]{2012A&A...547A...9P} Pons, J.~A., Vigan{\`o}, D., \& Geppert, U.\ 2012, \aap, 547, A9
\bibitem[Wu et al.(2003)]{2003A&A...409..641W} Wu, F., Xu, R.~X., \& Gil, J.\ 2003, \aap, 409, 641
\bibitem[Xu \& Qiao(2001)]{2001ApJ...561L..85X} Xu, R.~X., \& Qiao, G.~J.\ 2001, \apjl, 561, L85
\bibitem[Yue et al.(2007)]{2007AdSpR..40.1491Y} Yue, Y.~L., Xu, R.~X., \& Zhu, W.~W.\ 2007, Adv. in Space Res., 40, 1491
\bibitem[Zhang \& Xie(2011)]{2011ASPC..451..231Z} Zhang, S., \& Xie, Y.\ 2012a, in ASP Conf. Ser. 451, 4519th Pacific Rim Conference on Stellar Astrophysics, ed. Shengbang Qian (San Francisco, CA: ASP), 231
\bibitem[Zhang \& Xie(2012)]{2012ApJ...757..153Z} Zhang, S.-N., \& Xie, Y.\ 2012b, \apj, 757, 153 (Paper I)
\end{thebibliography}
\end{document}